\documentclass[pra,twocolumn,amsmath, superscriptaddress,notitlepage,showpacs]{revtex4-1}
\usepackage{amsmath,amsfonts,amssymb,amstext,amscd,amsthm,dsfont,bbm,hyperref,natbib,braket}
\usepackage[T1]{fontenc}
\usepackage{physics}
\usepackage{color}
\usepackage{soul,xcolor}
\usepackage{gensymb}
\usepackage{subcaption}
\hypersetup{
    colorlinks = true,
    citecolor=blue,
    linkcolor = red,
    anchorcolor = red,
    citecolor = blue,
    filecolor = red,
    pagecolor = red,
    urlcolor = blue,
    }
\usepackage[normalem]{ulem}
\usepackage{graphicx}
\usepackage{bbold}
\usepackage{braket}
\usepackage{placeins}

\begin{document}
\title{Noninvertibility and non-Markovianity of quantum dynamical maps}
		\author{Vinayak Jagadish}
	\email{vinayak.jagadish@uj.edu.pl}
	\affiliation{
Instytut Fizyki Teoretycznej, Uniwersytet Jagiello{\'n}ski, {\L}ojasiewicza 11, 30-348 Krak\'ow, Poland}
				\author{R. Srikanth}
\affiliation{Theoretical Sciences Division,
Poornaprajna Institute of Scientific Research (PPISR), 
Bidalur Post, Devanahalli, Bengaluru 562164, India}
\author{Francesco Petruccione}
	\affiliation{School for Data Science and Computational Thinking, Stellenbosch University, Stellenbosch 7600, South Africa}
	\affiliation{National Institute for Theoretical and Computational Sciences (NITheCS),  Stellenbosch 7600, South Africa}
\begin{abstract} 
We identify two broad types of noninvertibilities in quantum dynamical maps, one necessarily associated with CP indivisibility and one not so. We study the production of (non-)Markovian, invertible maps by the process of mixing noninvertible Pauli maps, and quantify the fraction of the same. The memory kernel perspective appears to be less transparent on the issue of invertibility than the approaches based on maps or master equations. Here we consider a related and potentially helpful issue: the identification of criteria of parameterized families of maps leading to the existence of a well-defined semigroup limit.
\end{abstract}
\maketitle  
\section{Introduction}
The major focus in the research on open quantum systems~\cite{petruccione} is on understanding the detrimental effects of noise, whereby information in the quantum system is leaked to the environment~\cite{haroche_exploring_2006}. The coherent evolution of quantum systems is impeded by the unavoidable coupling of quantum systems with the external environment. Combating this issue forms the basis of designing new quantum technologies. Conventionally, noise was treated to be Markovian or memoryless, but it is now well known that this assumption is false. Quantum non-Markovianity~\cite{breuer_colloquium:_2016,li_concepts_2017,de_vega_dynamics_2017}  is a beguiling theme in the study of open quantum systems, owing to both the fundamental aspects as well as its efficacy in the improvement of quantum information processing.

There are three equivalent ways in which reduced dynamics of the quantum system undergoing an open evolution is usually described: (a) The first way is by a generalization of the Schr{\"o}dinger equation, which is the time-local master equation given by
\begin{equation}
\begin{split}
 \mathcal{L}(t)[\rho]=&-\imath [H(t),\rho]\\&
+\sum_i \gamma_i (t)
\left(J_i(t)\rho J_i(t)^\dagger-\frac {1}{2}\{J_i(t)^\dagger J_i(t),\rho\}\right).
\label{gksl}
\end{split}
\end{equation}
Here, $H(t)$ is the effective Hamiltonian, $J_i(t)$'s are the Lindblad jump operators, and $\gamma_i (t)$ the decoherence rates, describing the time continuous dynamics of the system of interest. (b)The second way is by the family of solutions to Eq. (\ref{gksl}), which may be represented as a quantum dynamical map. This is a time-continuous family of completely positive (CP) and trace-preserving (TP)  linear time-dependent maps, $\{\mathcal{E} (t): t\geq 0, \mathcal{E}(0) = \mathbbm{1}\}$, acting on the bounded operators of the Hilbert space of the system of interest~\cite{sudarshan_stochastic_1961, Quanta77}. It follows from Eq. (\ref{gksl}) that the properties of the dynamical map are related to the time-local generator $\mathcal{L}(t)$~\cite{gorini_completely_1976} in the time-local master equation, $\dot{\mathcal{E}}(t) = \mathcal{L}(t)\mathcal{E}(t)$. (c) The third way to represent open system dynamics is via the memory kernel, introduced and discussed in Sec.~\ref{sec:kernel}. 

The divisibility of the dynamical map expresses the idea that the overall evolution from initial time $t_a$ to final time $t_b$ can be represented as a concatenation of intermediate propagators. This may be expressed by the following. Given arbitrary time instants $t_b\geq t_n \geq t_{n-1} \geq \cdots \geq t_2 \geq t_1 \geq t_a,$
\begin{align}
\mathcal{E}(t_b, t_a) &= V(t_b, t_n)V(t_n,t_{n-1}) \cdots V(t_2,t_1)\mathcal{E}(t_1, t_a)
\label{cpdivdef}
\end{align}
The map is CP divisible if, for all $t$, the intermediate propagators $V(t_m, t_{m-1})$ are CP. Correspondingly, the decay rates $\gamma_i (t)$ are positive at all times. Otherwise, the map is said to be CP indivisible. In Eq. (\ref{cpdivdef}), if for an instant of time, say $t = t^{*}$, the map $\mathcal{E}(t^{*},t_a)$ becomes noninvertible, then there are multiple initial states that are mapped to the same output state. At such instances, the propagator $V(t_b, t^{*}) =  \mathcal{E}(t_b, t_a) \mathcal{E}(t^{*} t_a )^{-1}$ is undefined. Such points $t^{*}$ are referred to as singular points~\cite{hou_singularity_2012} of the map. Even though the dynamical map could be momentarily noninvertible, the singularities are usually not pathological, in the sense that the reduced dynamics of the system of interest is regular~\cite{jagadish_measure2_2019, jagadish_initial_2021}.

As opposed to classical non-Markovianity, there have been a number of criteria proposed to witness and quantify quantum non-Markovianity~\cite{breuer_colloquium:_2016,li_concepts_2017,rivasreview}. The two major approaches to study quantum non-Markovianity,  are based  on the CP-indivisibility criterion~\cite{rivas_entanglement_2010, hall2010} and on the distinguishability  of  states~\cite{breuer_measure_2009}. According to the  Rivas-Huelga-Plenio (RHP) divisibility criterion~\cite{rivas_entanglement_2010}, a quantum dynamical map is Markovian if it is CP divisible at all instants of time.  Quantum non-Markovianity, therefore is indicated by a deviation from CP divisibility. Correspondingly, for a Markovian evolution, all the decay rates $\gamma_{i}(t)$ are non-negative, in the time-local master equation of the form given by Eq. (\ref{gksl}). The dynamical map is therefore non-Markovian, if any one of the decay rates turns negative at any instant of time. In the present paper, the characterization of non-Markovianity is done by analyzing the decay rates in the time-local master equation corresponding to the dynamical maps. According to the Breuer-Laine-Piilo (BLP) distinguishability or information backflow criterion~\cite{breuer_measure_2009}, a quantum dynamical map $\mathcal{E}(t)$ is said to be Markovian if it does not increase the distinguishability of two initial states $\rho_A$ and $\rho_B$, i.e., if $\Vert\mathcal{E}(t)(\rho_A) - \mathcal{E}(t)(\rho_B)\Vert \le \Vert\mathcal{E}(0)(\rho_A) - \mathcal{E}(0)(\rho_B)\vert\vert$, where $\Vert \cdot \Vert$ denotes the trace distance. The connection between noninvertibility and non-Markovianity was studied in~\cite{chruscinski2018}.

Recently, we studied the~\cite{jagadish_convex_2020} convex combinations of the set of Pauli semigroups and characterized the geometry of the resultant set along with providing a measure of the Markovian and non-Markovian regions. In~\cite{jagadish_nonqds_2020}, we considered the mixing of Pauli maps that are time-dependent Markovian. These results showed that the set of CP divisible Pauli maps is non convex, which was quantified via a non-zero measure of the resulting CP indivisible set. The non convexity of the set of generalized Pauli Markovian maps was studied in~\cite{siudzinska_jpa_2020}. An experimental realization of convex mixing of Pauli semigroups was realized on a NMR quantum information processor~\cite{gulati_2022}. Convex combinations of channels with singularities have also caught attention recently~\cite{siudzinska_markovian_2021,jagadish2022noninvertibility,siudzinska_jpa_2022,jagadish_measureinvert_2022}. Convex combinations of noninvertible qudit Pauli maps can produce a semigroup, as was reported in~\cite{siudzinska_markovian_2021}. Subsequently, it was shown that noninvertibility of the generalized Pauli input maps is necessary to create a semigroup~\cite{jagadish2022noninvertibility}. The fraction of the mixtures of noninvertible maps that produce (non)invertible maps was quantified in~\cite{jagadish_measureinvert_2022}.

In this article, we study the relationship between noninvertibility and non-Markovianity, focusing on certain issues pertaining to the classification, combinations, and representation of noninvertible dynamical maps. Specifically, we identify two different kinds of noninvertibilities based on the connection to non-Markovianity, and discuss their physical origin. We then focus on the convex combinations of noninvertible Pauli maps and quantify the measure of non-Markovian maps within the set of invertible maps obtained by mixing. We also briefly look into certain aspects of memory kernels corresponding to noninvertible maps.

The paper is organized as follows. In Sec.~\ref{sec:noninvert}, we point out how the question of connection between noninvertibility and non-Markovianity naturally leads to a classification of noninvertibility into two types, based on whether or not it is accompanied by non-Markovianity. Next we consider another aspect of this connection: namely, the production of non-Markovian invertible maps by mixing noninvertible maps. Invertible maps can be obtained by mixing noninvertible Pauli maps, and the fraction of non-Markovian maps therein are derived in Sec.~\ref{sec:paulisec}. To all appearances, the memory kernel representation is less transparent on questions pertaining to noninvertibility, much less its connection to non-Markovianty, than the dynamical map or master equation representation. As an initial effort in connecting that situation, the memory kernel perspective of families of maps with a semigroup limit is discussed in Section \ref{sec:kernel}. Finally, we conclude with discussions in Sec.~ref{conclusion}.

\section{Types of noninvertibilities \label{sec:noninvert}}
In the many examples of (non)invertibility cited above, typically the noninvertibility goes hand in hand with non-Markovianity such that the onset or disappearance of non-Markovianity coincides with the occurrence of a singularity. This prompts the question of whether this phenomenon is generic and, more generally, the question of relationship between noninvertibility and non-Markovianity.

A singularity corresponds to a point $\rho^{\star}$ in the  state space of the system of interest where multiple evolutionary tracks (originating from different initial states) converge in finite time. By the continuity argument, either these tracks must diverge or $\rho^{\star}$ is a fixed point. These two possibilities correspond to two distinct types of singularities.

The latter possibility corresponds to a ``type I'' noninvertibility, which may be encountered in the case of decoherence due to the  interaction of the system with an infinite bath.  The former possibility corresponds to a ``type II'' noninvertibility, discussed later below. Type I noninvertibility need not be associated with non-Markovianity. By contrast, typically, type II noninvertibility marks the onset of non-Markovianity, indicated by a temporarily negative decoherence rate in the master equation.

Consider the qubit dephasing map
\begin{equation}
\mathcal{E}_3^q(\rho) = [1-q(t)]\rho + q(t)\sigma_3\rho \sigma_3,
\label{eq:deph}
\end{equation}
 where $q(t)$ is the decoherence function and $\sigma_3$ is the Pauli matrix. 

The corresponding time-local master equation of the map is
\begin{equation}
\label{megen}
\Dot{\rho} = \mathcal{L}(t)(\rho) \equiv \gamma_3 (t)(\sigma_3\rho\sigma_3-\rho),
\end{equation}
The decay rate $\gamma_3 (t)$ turns out to be
\begin{equation}
\gamma_3 (t) = -\frac{\dot{\lambda}(t)}{\lambda(t)}=\frac{2 \dot{q}(t)}{1-2 q(t)}, 
\label{ratesdecay1}
\end{equation}
where it is related to the eigenvalue of the map $\lambda(t) = 2q(t)-1$.

\subsection{Type I noninvertibility \label{sec:I}}
As an example, we consider the dephasing map of Eq. (\ref{eq:deph}), with the decoherence function $q(t)$ chosen such that $q(t)$ monotonically increases up to a finite time $t^{\star}$, such that $q(t^{\star})=\frac{1}{2}$, and it remains constant thereafter. For example:
\begin{equation}
    q(t) = f(t)[1-\Theta(t-t^{\star})] + \frac{1}{2}\Theta(t-t^{\star}),
    \label{eq:heavyp}
\end{equation}
where $f(t)$ is any monotonically increasing function such that $f(0) = 0, f(t^{\star})=\frac{1}{2}$ and $\Theta(t)$ is the Heaviside function. Thus, each input map is maximally dephasing and thereby noninvertible (with all ``azimuthal'' points on the Bloch ball mapped to a single point on the respective axis). 

This yields a rate, in view of Eq. (\ref{ratesdecay1}), given by
\begin{equation}
    \gamma_3 (t) = \frac{\Dot{f}[1-\Theta(t-t^{{\star}})] + (0.5 - f)\delta(t^{{\star}})}{1-2q(t)}
    \label{eq:thetarate}
\end{equation}
Thus the decay rate remains positive until the singular instance $t^{{\star}}$, where it attains the indeterminate value $\frac{0}{0}$ and remains so for $t> t^{{\star}}$. The situation can essentially be understood as a consequence of the system reaching a fixed point at $t^{{\star}}$. Referring to Eq. (\ref{megen}), the $\Dot{\rho}$ in the LHS and the RHS term $(\sigma_3\rho\sigma_3-\rho)$ both identically vanish, making $\gamma_3(t)$ indeterminate for $t\ge t^{{\star}}$. Furthermore, starting from the fixed point, the eigenvalues of the map Eq. (\ref{eq:deph}) vanish for all $t \ge t^{{\star}}$.

The evolution is Markovian in the time $t \le t^{\star}$ since the rates are all positive. The evolution also remains Markovian for the subsequent time, because the system remains ensconced in the fixed point, implying that the intermediate map is the identity map, which is CP. Thus, a type I noninvertibility need not be associated with non-Markovianity.
On the other hand, type I noninvertibility, if it holds, must arise from an infinite dimensional environment. This is because a finite dimensional system is typically associated with finite-time recoherence, given the overall system-bath reversibility of the interaction. To this end, consider the closed system consisting of system $S$ and a finite environment $E$, and interacting via a Hamiltonian $H_{SE}$. Started in an arbitrary state $\ket{\Psi}_{SE}$, the state $e^{-i\hbar H_{SE}t}\ket{\Psi}_{SE}$ rotates the system $S$ away from and back to its initial state in a time scale about $\hbar/|H_{SE}|$, where $|H_{SE}|$ denotes the largest eigenvalue of $H_{SE}$. This rules out that a fixed point can be attained by a system interacting with a finite environment.

\subsection{Type II noninvertibility \label{sec:II}}
The other possibility of noninvertibility, called ``type II'' noninvertibility, is the more familiar type. In this case, the Markovian evolutionary tracks that converge towards the singular point now continuously diverge from each other. This part of the evolution produces recoherence and/or information back flow, corresponding to non-Markovianity that is CP indivisible or P indivisible. For this same reason, it seems safe to rule out a Markovian evolution, not to mention semigroup, if the bath is finite dimensional.
Characteristically, one of the canonical Lindblad rates turns negative after the instant $t^\ast$ of the singularity. Thus, noninvertibility signals non-Markovianity. In this case, a negative rate indeed entails non-Markovianity, as in the case of invertible maps~\cite{hall2010}.

Type II noninvertibility 
is encountered in the case of decoherence due to the  interaction of the system with a finite bath, which allows the possibility of information back-flow or recoherence in finite time. As an example, let $q(t)$ be chosen as $\frac{1}{2} [1-\cos (\omega t  )]$. This can arise due to the interaction of a qubit  with a single-qubit environment ($E$) initialized in the  state $\frac{1}{\sqrt{2}}(\ket{0}+\ket{1})$, with the  Hamiltonian 
\begin{equation}
\label{Hamiltonian}
H = \frac{\omega}{2}\sigma_{3}\otimes\sigma_{3}^{E}.
\end{equation}
The corresponding time-local master equation is of the form given by Eq.~(\ref{megen}), with the decay rate
\begin{equation}
\label{decayratetan}
\gamma_3 (t) = \frac{\omega  \tan ( \omega t )}{2 }.
\end{equation} 
In this case, the map must flip the sign of a decay rate after a singularity. Indeed, the first instance $t^\star ~(< \infty)$ of noninvertibility marks the onset of non-Markovianity. This corresponds to the situation where multiple evolutionary tracks diverge, having converged at the singularity. The point of convergence leads to an instantaneous indistinguishability of the corresponding convergent states. By continuity these tracks diverge, as noted. The diverging paths correspond to increasing distinguishability, and hence non-Markovianity. 

Note that this stands in contrast to case of type I noninvertibility, which corresponds to a fixed point, so that there is no divergence of evolutionary tracks. Here, we could also consider the type I noninvertibility as one that is obtained from type II in the limit that the size of the environment increases towards infinity, and correspondingly the backflow time period rises to the Poincar\'e recurrence time~\cite{pottier}.

It must be noted that a finite-dimensional environment does not always lead to a non-Markovian evolution. If the system qubit interacts with a single-qubit environment initialized in the state $\ket{0}$ under the same Hamiltonian, given by Eq.~(\ref{Hamiltonian}), then the reduced dynamics on the system is unitary, essentially because the  state is an eigenstate of the operator $\sigma_3^{E}$. This example also shows that a finite-dimensional environment need not always lead to noninvertibility.

Suppose we force the decay rate to be positive beyond the singularity, by requiring it to be, say, $\propto \tan^2(\omega t)$. This would ensure that the rate does not flip sign beyond the singular point, but using Eq.~(\ref{ratesdecay1}), one can see that the differential equation leads to an invalid probability function for $q(t)$. This can also be understood by noting that just after the singularity, the converging evolutions diverge, indicating that a sign flip of the decay rate to negative is necessary.

A shared feature of both types of noninvertibility is that the singularity is a point of entanglement sudden death. Given the maximally entangled state $\ket{\Phi^+} \equiv \frac{1}{\sqrt{2}}(\ket{00} + \ket{11})$, for the map given by Eq. (\ref{eq:deph}), we find that $(\mathcal{E}^{1/2}_3 \otimes \mathbb{1})\ket{\Phi^+}$ is a separable state, at the singularity of $\mathcal{L}$ in Eq. (\ref{megen}). In other words, the Choi-Jamio{\l}kowski matrix corresponding to either type of map is a separable state. The difference is that with type II, there is a subsequent revival of entanglement, but not with type I.

\section{Convex Combinations of Type-II noninvertible Pauli dynamical maps and measure of the non-Markovian set}
\label{sec:paulisec}
In~\cite{jagadish_measureinvert_2022}, the fraction of invertible maps obtained by mixing noninvertible ones was evaluated. These correspond to the type II noninvertible case.  Here, we quantify the fraction of non-Markovian maps  in the case where the resultant is an invertible map. 
To this end, let us consider the convex combination of the three Pauli dynamical maps, each mixed in proportions of $x_i$ as
\begin{equation}
\label{outputmappauli}
\tilde{\mathcal{E}}(t) = \sum_{i=1}^{3} x_{i} \mathcal{E}_i (t),  \quad \big(x_i >0, \sum_i x_i =1\big),
\end{equation}
with
\begin{equation}
 \label{paulichanndef}
\mathcal{E}_i (t)[\rho]=[1-q(t)]\rho + q(t)\sigma_i\rho \sigma_i,\thinspace i= 1,2,3,
\end{equation} 
The three $\mathcal{E}_i (t)$ shall be called input maps and $\tilde{\mathcal{E}}(t)$  the output map. Each of the maps are characterized by the same decoherence function, 
\begin{equation}
\label{decohfunc}
q(t) = \frac{1-e^{-jt}}{m}, \quad m\geq 1, j >0.
\end{equation}

The eigenvalue relation for the output map $\tilde{\mathcal{E}}$ [Eq. (\ref{outputmappauli})] reads
\begin{equation}
\label{eigpauli}
\tilde{\mathcal{E}}(t)[\sigma_i] = \lambda_i (t) \sigma_i,
\end{equation}
with
\begin{equation}
\lambda_i(t) =1- 2(1-x_i)q(t).
\label{eq:lambdaz}
\end{equation} 
For $q(t)$ given by Eq. (\ref{decohfunc}), the eigenvalues $\lambda_i (t)$ become singular at
\begin{equation}
t^{*}=\frac{1}{j}\ln \Big[\frac{2(1-x_i)}{2(1-x_i)-m}\Big].
\label{eq:tstar}
\end{equation}
For $m \ge 2$, the input and output maps are all invertible. 
From Eq. (\ref{eq:tstar}), for the output map to be invertible, 
\begin{equation}
x_i \ge 1-\frac{m}{2}.
\label{eq:xi2}
\end{equation}
For $m < \frac{4}{3}$, one obtains the constraint $x_i < \frac{1}{3}$, which cannot be satisfied by all three maps (since in that case we would have $\sum_i x_i < 1$). For noninvertible inputs, if $m< \frac{4}{3}$, the output map is always noninvertible. For input maps in the noninvertible range ($2 > m > \frac{4}{3}$), the invertibility of the output map depends on the mixing coefficients $x_i$, as seen from Eq. (\ref{eq:xi2}). The fraction  $\Delta_{\rm invert}$ of invertible maps obtained by mixing noninvertible maps in the range $m \in (4/3,2)$ was evaluated in~\cite{jagadish_measureinvert_2022} as $\frac{(4-3 m)^2}{8}$. The fraction of invertible maps falls monotonically through the range $[1,0]$ for varying $m$ in $(\frac{4}{3},2)$. 

Within the fraction of invertible maps obtained by mixing noninvertible maps, we now quantify the fraction of Markovian and non-Markovian regions. To this end, one needs the the time-local generator for the output map $\tilde{\mathcal{E}}(t)$, which can be evaluated to be
 \begin{equation}
\label{meqgen}
\mathcal{L}(t) \rho = \sum_{i=1}^{3}\gamma_{i}(t)(\sigma_i\rho\sigma_i-\rho),
\end{equation} with the decay rates 
\begin{equation}
\gamma_i(t) = \left(-\frac{1-x_i}{1-2 (1-x_i) q(t)} + \sum_{j\neq i} \frac{1-x_j}{1-2 (1-x_j) q(t)}\right)\frac{\dot{q}(t)}{2}.
\label{ratesdecaythree}
\end{equation}
Let us now discuss the geometry and fraction of the Markovian and non-Markovian maps in the parameter space of $(x_1,x_2)$.
From the functional form of the decay rates in Eq. (\ref{ratesdecaythree}), we can see that they have the structure-
\begin{equation}
\gamma_i(t) = -f(x_i,q(t))+  \sum_{j\neq i} f(x_j,q(t)),
\label{eq:3form}
\end{equation} 
where $f(x_i,q(t))\ge0$ for all $q[t] \in [0,\frac{1}{m})$. The sum of any two decay rates, $\gamma_{i}(t) + \gamma_{j}(t), where i,j = 1,2,3 and i \neq j$ is always positive, even though an individual rate may be negative. For instance, $\gamma_2(t) + \gamma_3(t) = 2f(x_1, q(t)) \ge 0$.  This indicates that the dynamics obtained by convex combination is P divisible and hence Markovian according to the BLP distinguishability criterion.

From Eq. (\ref{ratesdecaythree}), one can see that if a given rate, say $\gamma_2(t)$, turns negative at $q=q_0 \le \frac{1}{m}$, then it remains negative throughout the remaining range of $[q_0,\frac{1}{m}]$.  To find the set of all pairs $(x_1,x_2)$ such that $\gamma_2(t)\le0$ at $q=\frac{1}{m}$, one solves the equation $\gamma_2(x_1,x_2,\frac{1}{m},t) = 0$ for a constraint on $x_1$ and $x_2$. This yields
\begin{equation}
x_1^\pm(x_2) \equiv \frac{1}{2} 
 \left(\pm\frac{\eta(n,x_2)}{x_2+(m-1)}-x_2+1\right),
\label{eq:xmaxmin}
\end{equation}
where
\begin{widetext}
\begin{align}
\eta(m,x_2) &= \left[(-m+x_2+1) (m+x_2-1) \left(\mu_m^+-x_2\right)\left(\mu_m^--x_2\right)\right]^{\frac{1}{2}},\nonumber\\
\mu_{m}^{\pm} & = \pm\sqrt{m^2+1}-m.
\end{align} 
\end{widetext}
It can be seen that the values $x_1^\pm(x_2)$ are real only in the range $x_2 \in [1-\frac{m}{2}, \mu_m^+]$. Eq. (\ref{eq:xmaxmin}) implies that $\gamma_2(t)<0$ for the values $x_1 \in (x_1^-(x_2), x_1^+(x_2))$ where any given $x_2$ is in the allowed range. Corresponding to these points $(x_1, x_2) $, we determine the region $\mathfrak{R}_{x_2}$ which yield a negative $\gamma_2(t)$:
\begin{equation}
|\mathfrak{R}_{x_2}| = \frac{1}{\Delta_{\rm invert}}  \int_{x_2=1-\frac{m}{2}}^{\mu_{m}^+} \left[x_1^+(x_2) - x_1^-(x_2)\right]dx_2.
\label{eq:mu1}
\end{equation} 
Eq. (\ref{ratesdecaythree}) implies that at most only one of the three rates can be negative. Therefore, the regions $\mathfrak{R}_{x_1}, \mathfrak{R}_{x_2}$ and $\mathfrak{R}_{x_3}$, respectively, of points $(x_1,x_2,x_3)$, where $\gamma_1(t), \gamma_2(t)$, and $\gamma_3(t)$ can assume negative values within the time range $q \in [0, \frac{1}{n}]$, are non overlapping. Therefore, the fraction $|\mathcal{F}|$ of the set  of all non-Markovian maps within the region of invertible maps obtained by mixing noninvertible Pauli maps is $|\mathcal{F}| = 3|\mathfrak{R}_{x_2}|$. 

As an example, consider the case of $m=5/3$. Following the calculations as discussed above, one obtains
\begin{widetext}
\begin{eqnarray}
|\mathcal{F}| &=& 3 \times 8  \int_{x_2=1/6}^{\frac{1}{3}(-5+\sqrt{34})}\frac{\sqrt{9 x_2^4+30 x_2^3-13 x_2^2-\frac{40 x_2}{3}+4}}{3 x_2+2}dx_2 \nonumber\\
&\approx&0.826203.
\end{eqnarray} 
\end{widetext}
The factor 8 is due to the fraction of invertible maps, which, for $m=5/3$, is 1/8.

\begin{figure}[ht!]
    \centering  
\includegraphics[width=0.5\textwidth]{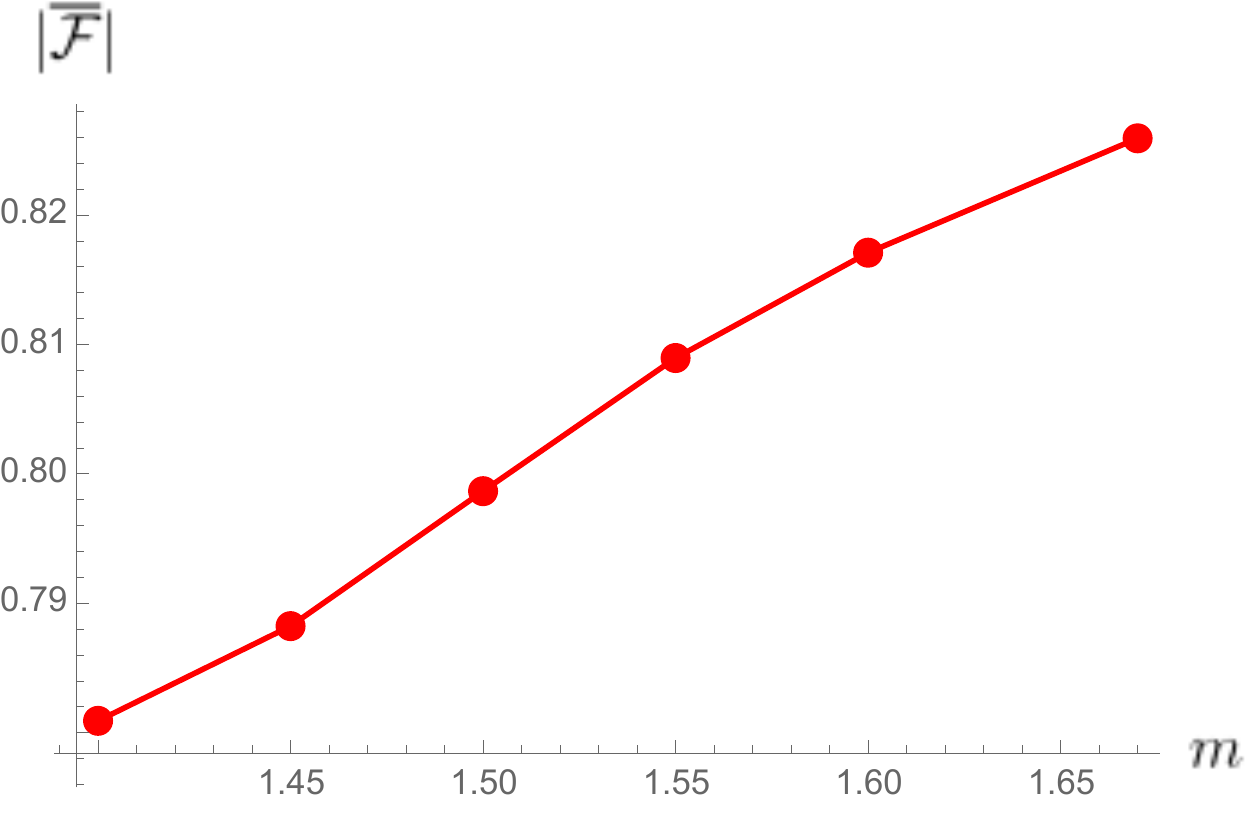}
	\caption{(Color online) Plot of the fraction $|\overline{\mathcal{F}}|$ of non-Markovian maps with varying $m$. }
	\label{fig:measure}
\end{figure}

A plot of the fraction $|\mathcal{F}|$  of non-Markovian maps with varying $m$  is shown in Fig. \ref{fig:measure}. 

\begin{figure}[ht!]
    \centering  
\includegraphics[width=0.55\textwidth]{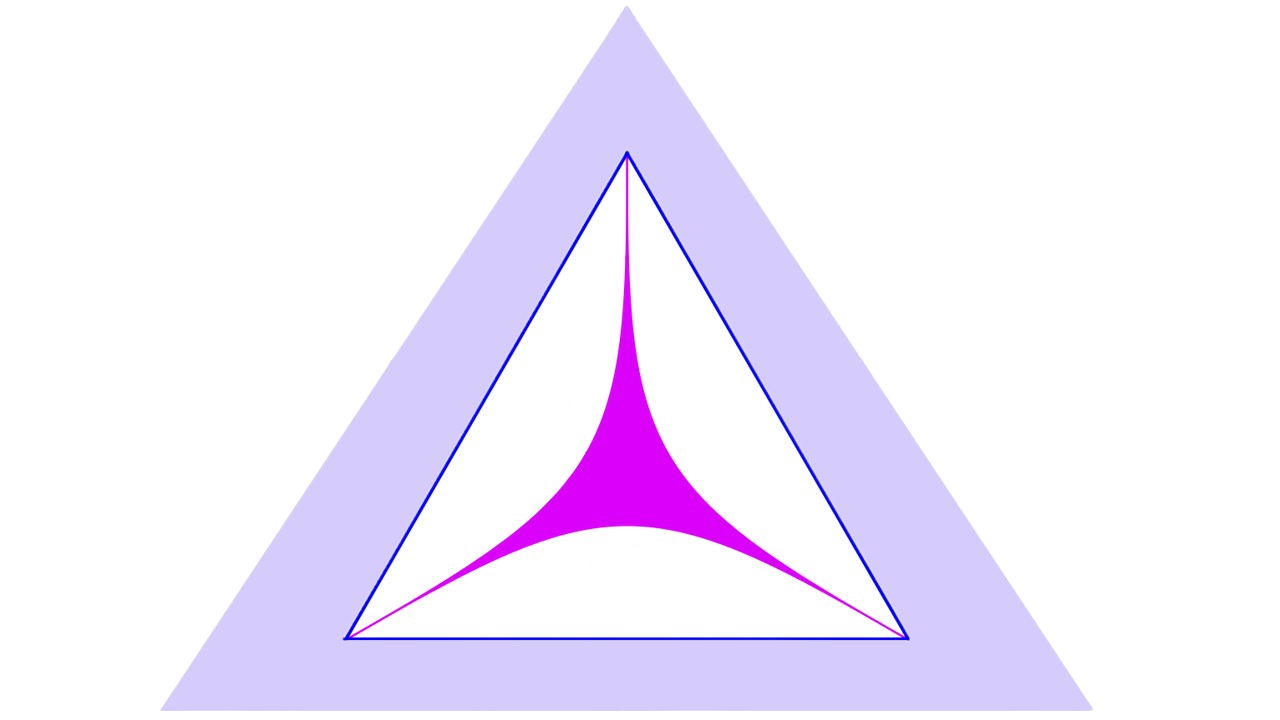}
	\caption{(Color online) The outer equilateral triangle is the Pauli simplex with vertices given by noninvertible Pauli maps with $m=\frac{5}{3}$. The position within the triangle represents the relative proportion with which the three maps are mixed. The inner  equilateral triangle is a simplex corresponding to a restriction to mixing coefficients $\frac{1}{6} \le x_i \le 1$ and represents invertible output maps. Thus the space between the two equilateral triangles corresponds to noninvertible output maps. 
  The squeezed (shaded) triangle that sits tightly within the inner equilateral triangle corresponds to mixtures that yield a Markovian dynamical map.}
	\label{fig:nmsimplex}
\end{figure}

A diagrammatic representation of the set of Markovian and non-Markovian maps as per our above analysis is given  in Fig.~\ref{fig:nmsimplex}. Here the outer equilateral triangle is a Pauli simplex with the vertices representing the three Pauli maps with $m=\frac{5}{3}$. The regions $\mathfrak{R}_{x_1}$, $\mathfrak{R}_{x_2}$ and $\mathfrak{R}_{x_3}$ in the $(x_1,x_2,x_3)$ parameter space, where $\gamma_1$, $\gamma_2$, or $\gamma_3$ turn negative, will be non-overlapping. The squeezed triangle  represents the non-convex Markovian region. 

\section{Parametric families of map with or without a semigroup limit \label{sec:kernel}}
There exists an alternate representation of the dynamics of open quantum systems in terms of an integro-differential equation, which is referred to as the memory kernel master equation.
\begin{equation}
\dot{\mathcal{E}}(t)= \int_{t_0}^{t} d\tau K(t,\tau) \mathcal{E}(\tau),
\label{nzeqndef}
\end{equation}
where $K(t,\tau)$ is known as the memory kernel. For a given dynamical map, one can in principle find the associated time-local generator as well as the memory kernel. However, the 
necessary and sufficient conditions for $K(t,\tau)$ that generate a legitimate (completely positive and trace-preserving) dynamical map are still not completely known.
Using the property of Laplace transforms, one can write the same  as
\begin{equation}
s\hat{\mathcal{E}}(s) - \mathcal{E}(0) = \hat{K} (s) \hat{\mathcal{E}}(s),
\label{nzeqnlt}
\end{equation}
and therefore
\begin{equation}
 \hat{K} (s) = s \mathbbm{1} - \hat{\mathcal{E}}^{-1} (s),
\end{equation}
with $\hat{K} (s)$ being the corresponding kernel after Laplace transform, defined as $\hat{f}(s) = \int_{0}^{\infty}e^{-st}f(t)dt$.

As a simple example, consider the Pauli dephasing map, given by Eq. (\ref{eq:deph}), with $q(t)$ to be $\frac{1}{2} [1-\cos (\omega t)]$. Clearly, the map is noninvertible, with the singularities at $\pm n\pi/2$.
The corresponding memory kernel can be evaluated to be
\begin{equation}
K(t)(\rho) =  \frac{\omega ^2}{2}\Theta(t)(\sigma_3\rho\sigma_3-\rho).
\end{equation}
Surprisingly, the kernel is regular as opposed to the dynamical map or the time-local master equation.
Note that the eigenvalue $\kappa(t)$ of the operator $K(t)$  is $\omega ^2 \Theta(t)$, which implies that
$\lambda(t) = \cos (\omega t)$.

Consider the Pauli dephasing map,  given by Eq. (\ref{eq:deph}), with $q(t)$ as in Eq.~(\ref{decohfunc}).
The corresponding time-local master equation is of the form given by Eq.~(\ref{megen}) with the decay rate, 
\begin{equation}
\gamma_3 (t) = \frac{c}{(n-2)e^{c t}+2}.
\label{ratesdecay}
\end{equation}

Switching to the memory kernel, in the time-domain after the inverse Laplace transform, one obtains 
\begin{equation}
K(t)(\rho) =  \Big(\frac{ c \delta (t)}{n}-\frac{ c^2 (n-2) e^{-\frac{c (n-2) t}{n}}}{n^2}\Big)\Big(\sigma_3\rho\sigma_3-\rho\Big).
\end{equation}
Clearly, for $n=2$, the memory kernel reduces to the $\delta$ distribution, indicative of a semigroup.

One should also note that the Markovian limit is not always recovered in the memory kernel description. To this end,
we now  discuss  a  dephasing model motivated by the random telegraph
noise  (RTN).  Let $q(t)$ in Eq. (\ref{eq:deph}) be

\begin{eqnarray}
q(t) = e^{-\alpha t}\left[\textmd{cos}\left(\omega\alpha t \right)+
\frac{\textmd{sin}\left(\omega\alpha t\right)}{\omega}\right], \thinspace \alpha,\omega\geq0.
\label{kernal}
\end{eqnarray}

The time-local master equation for the map is of the form given by Eq. (\ref{megen}), with the decay rate
\begin{equation}
\label{megenrtn}
\gamma_3(t) = \frac{\alpha  \left[1+\omega ^2\right]}{2[1+\omega  \cot (\alpha  t \omega )]}.
\end{equation}

The associated memory kernel turns out to be a purely non-local one (with no term proportional to the $\delta$ distribution), which is
\begin{equation}
K(t)(\rho) = \frac{\gamma ^2 }{2}\left(\omega ^2+1\right) e^{-2 \gamma  t} (\sigma_3\rho\sigma_3-\rho).
\end{equation}

The above examples show that noninvertibility necessarily contains a nonlocal element in the kernel. This raises the question of what feature of the map corresponds to a purely nonlocal kernel. The above exercise suggests, quite straightforwardly, that there should be no parametric limit in which the map goes to the exponential form characteristic of a semigroup. This is readily verified in Eq. (\ref{kernal}): in the limit $\omega\rightarrow 0$, the sine term blocks a purely exponential form (noting that $\lim_{\omega\rightarrow 0} \frac{\sin(\omega\alpha t)}{\omega} = \alpha t$), whereas in the limit $\omega\rightarrow\infty$, the cosine term is the culprit. 

 Consider a modified RTN, with 
\begin{eqnarray}
q(t) = e^{-\alpha t}\left[\textmd{cos}\left(\omega\alpha t \right)+
{\textmd{sin}\left(\omega\alpha t\right)}\right], \thinspace \alpha,\omega\geq0,
\label{eq:kernel}
\end{eqnarray} 
in place of Eq. (\ref{kernal}). One directly verifies that in this case, the limit $\omega \rightarrow 0$ indeed leads to the exponential form. The corresponding memory kernel turns out to be
\begin{equation}
K(t)(\rho) = [\gamma ^2 \omega ^2 e^{-\gamma t (1+\omega)}-\frac{\gamma}{2}(\omega-1)\delta(t)](\sigma_3\rho\sigma_3-\rho).
\end{equation}
The presence of the local term may be noted, which ensures that the kernel is purely local in the limit $\omega\rightarrow0$.

\section{ Discussions and Conclusions
\label{conclusion}}

In this article, we studied various aspects of noninvertible dynamical maps. A quantum dynamical map can be noninvertible either because the system equilibrates with an infinite bath in finite time or it oscillates through a singularity via interaction with a finite environment. In the latter case, typically there is a sign flip of one or more canonical decoherence rates, and the map is necessarily non-Markovian. In the former case, the evolution, though noninvertible, is Markovian, and the decay rates are indeterminate after the equilibrium point. We then considered convex mixtures of noninvertible maps and calculated the fraction of non-Markovian maps within the set of invertible maps obtained by mixing. We also characterized the geometry and measure of the set for a particular case. We also looked into the memory kernel description of noninvertible maps through various examples.

What would be the ramifications of time-coarse-graining used to derive microscopic master equations describing open dynamics on noninvertible maps? Can one devise strategies for quantum error mitigation in the presence of noise with singularities? The complementary aspects of the time-local generator $\mathcal{L}(t)$ and the memory kernel $K(t)$ need to be understood in more detail. These are several questions for which we seek answers.
\color{black}

\acknowledgements

V.J. acknowledges financial support by the Foundation for Polish Science
through TEAM-NET project (Contract No. POIR.04.04.00-00-17C1/18-00). R.S.   acknowledges the support of Department of Science and Technology (DST), India, through Grant  No.CRG/2022/008345,
the Interdisciplinary Cyber Physical Systems (ICPS) programme Grant No. DST/ICPS/QuST/Theme-1/2019/14. F.P. acknowledges support of the NICIS (National Integrated Cyber Infrastructure System) e-research grant QICSA and of the South African Quantum Technology Initiative (SA QuTI).

\end{document}